\documentclass[reprint,amsmath,amssymb,aps,superscriptaddress,twocolumn,float]{iopart}
\usepackage{graphicx}
\usepackage{graphics}
\usepackage{dcolumn}
\usepackage{bm}
\usepackage{amsfonts}
\usepackage{amssymb}
\usepackage{float}
\usepackage{xcolor}
\newcommand{\angstrom}{\textup{\AA}}

\begin{document}
	
\title[]{Dirac Nodal Line and Rashba Splitting Surface States in Nonsymmorphic ZrGeTe}

\author{Yun Yen$^1$, Cheng-Li Chiu$^2$, Ping-Hui Lin$^3$,\\ Raman Sankar$^{2,4}$, Fangcheng Chou$^4$, Tien-Ming Chuang$^2$, and Guang-Yu Guo$^{1,5}$}

\address{$^1$ Department of Physics and Center for Theoretical Physics, National Taiwan University, Taipei 10617, Taiwan}
\address{$^2$ Institute of Physics, Academia Sinica, Nankang, Taipei 11529, Taiwan}
\address{$^3$ National Synchrotron Radiation Research Center (NSRRC), Hsinchu 30076, Taiwan}
\address{$^4$ Center for Condensed Matter Sciences, National Taiwan University, Taipei 10617, Taiwan}
\address{$^5$ Physics Division, National Center for Theoretical Sciences, Hsinchu 30013, Taiwan}

\begin{abstract}

Dirac semimetals (DSMs) are three dimensional analogue to graphene with symmety enforced bulk Dirac nodes. Among various DSMs, ZrSiS has been attracting more interests recently, due to its three dimensional Dirac nodal line protected by the nonsymmorphic symmetry. It actually belongs to a large family of isostructural compounds with unique quantum phenomenon. Here we present a comprehensive study of the first principle calculation, angle-resolved photoemission spectroscopy (ARPES) measurements, and scanning tunneling microscope (STM) experiments on ZrGeTe, a member of the ZrSiS family with stronger spin-orbit coupling (SOC). Our band structure calculation shows the existence of floating gapless surface states at $\bar{X}$ with Rashba splitted helical spin texture, which are confirmed by our ARPES measurements. We also perform quasiparticle scattering interference (QPI) imaging and find several q-vectors, with two Umklapp scattering vectors not observed in other family compounds. All the q-vectors can be identified with joint density of states (JDOS) simulation. Our results demonstrate the interesting electronic structure of ZrGeTe and might benefit the potential application by utilizing its exotic quantum states in the future.

\end{abstract}

\section*{Introduction}

Dirac semimetals (DSMs), which belong to one type of topological semimetals (TSMs), are three dimensional analogue to graphene and characterized by linear dispersed bulk bands described by the Dirac Hamiltonian \cite{burkov2011topological,burkov2016topological}. The Dirac bands can form either nodal points or nodal lines in the band structure. Usually these crossings can be gapped out by the spin-orbit coupling (SOC). However, with the existence of nonsymorphic symmetry, Dirac points or Dirac nodal-lines can be robust under the effect of SOC. This is first predicted by Kane and Young \cite{young2015dirac}, and further confirmed in the angle-resolved photoemission spectroscopy (ARPES) study on Dirac nodal-line semimetal ZrSiS \cite{schoop2016dirac,neupane2016observation,chen2017dirac,hosen2017tune} with nonsymmorphic space group \textit{P}4/\textit{nmm}. Ever since, several studies on ZrSiS family compounds reveal their similar electronic structures \cite{topp2016te,hosen2017tune,hosen2018observation,nakamura2019evidence} with some exotic quantum phenomenom. The glide mirror symmetry preserves the gapless Dirac nodes pinned at time reversal invariant momentas (TRIMs) and line nodes on the Brillounin zone (BZ) boundary \cite{young2015dirac,chen2017dirac}. The unique diamond-shaped Fermi surfaces lead to highly anisotropic magnetoresistance \cite{ali2016butterfly,su2018surface} and U-shaped optical conductivity \cite{allah2019optical}. The "floating "surface states due to the broken glide symmetry on the terminated plane in such systems are also intriguing \cite{topp2017float}. For example, HfSiS exhibits Rashba splitting in linearly dispersed surface bands \cite{chen2017dirac}, while ZrGeTe has gapless surface bands in topological crystalline insulator (TCI) phases \cite{hosen2018observation}. The ARPES and quasiparticle scattering interference (QPI) imaging allow us to unravel the nature of these surface states experimentally. ZrSiS shows strong band-selective scattering due to the magnetic quantum number conservation \cite{butler2017quasiparticle}. The QPI q-vectors in such systems show common features, including scattering between diamond-shaped bands and the anomalous Umklapp scattering between surface states at $\overline{X}$ \cite{su2018surface,butler2017quasiparticle,lodge2017observation,zhen2018nonsym}. The latter is discussed recently with the nonsymmorphic effect in ZrSiSe \cite{zhen2018nonsym}.

In this paper, we present a comprehensive study on ZrGeTe, a compound in the ZrSiS family with stronger SOC strength, by density functional theory (DFT) calculations, ARPES measurements, and STM imaging. Our bulk calculation shows that ZrGeTe has nonsymmorphic protected Dirac nodes at TRIMs and line nodes on the zone boundary, which are common features in the ZrSiS family. In addition, we also find a type-II Dirac point near $\overline{\Gamma}$ with negligible gap opened by SOC. Furthermore, the slab calculation shows that the linearly dispersed floating gapless surface band at $\overline{X}$ exhibits Rashba splitting with helical in-plane spin texture. The ARPES measurements support the calculated band structures with detected Rashba splitting. From our QPI imaging data, we find five prominent dispersing \textbf{q}-vectors including hallmark scattering process in the ZrSiS family \cite{su2018surface,butler2017quasiparticle,lodge2017observation,zhen2018nonsym} as well as several strong Umklapp scattering. Our results show that ZrGeTe and the related compounds are valuable platforms to study interplay of several quantum phenomena, including symmetry protected Dirac line nodes, Rashba effect, and gapless surface states in TCI phases.

\section*{First principle calculation}

In this work, the ZrGeTe single crystal is grown by the chemical vapor transport technique with lattice constant a = 3.866 $\angstrom$ and c = 8.599 $\angstrom$ determined by the X-ray diffraction (Details regarding crystal growth are discussed in the appendix). The crystal structure is shown in Fig. 1(a). ZrGeTe crystallizes in tetragonal structure \textit{P}4/\textit{nmm} space group, which possesses a nonsymmorphic glide mirror operation $(M_z|\frac{1}{2}\frac{1}{2}0)$. The calculated bulk electronic band structure without and with the SOC are shown in Fig. 1(b) and Fig. 1(c), respectively. The Dirac nodes pinned at high symmetry points (X, M, R, and A) marked with red circles are protected by the glide mirror symmetry \cite{young2015dirac,chen2017dirac}. They cannot be gapped by the SOC and thus form protected Dirac line nodes along A-M and X-R marked with red boxes \cite{chen2017dirac}. On the other hand, high symmetry lines $\Gamma$-X, X-M, Z-R, and R-A all belong to little group $C_{2v}$, which only has one irreducible representation under the effect of the SOC. As a result, all the crossings along these lines must be gapped by the SOC due to the band hybridization. These $C_{2v}$ SOC-gapped Dirac nodes are marked in blue circles. Interestingly, there is one type-II Dirac point near $\Gamma$ without considering the spin degree of freedom, and the crossing is ensured by the band inversion of $A_2$ and $B_2$ bands shown in the green inset of Fig. 1(b). In principle, the point should be gapped by the SOC since it is in the $C_{2v}$ region, while our calculation shows that the gap is smaller than 10$^{-4}$ eV, which is close to the computational precision. 

We also perform 9 layers slab calculation in order to study the surface states. The slab band structure with ratio of surface component is presented in Fig. 1(d). The floating surface band observed in other isostructural compounds \cite{topp2017float,zhen2018nonsym} is also seen here, dispersing linearly from -0.3 eV to 1.0 eV along $\overline{XM}$. Interestingly, the surface band forms a gapless node with lower bands at X point. By a recent symmetry indicator analysis \cite{song2018quant,zhang2019catalogue,verg2019catalogue}, which shows that $Z_{2,2,2,4}$= (0012) for ZrGeTe and corresponds to the nontrivial topological crystalline insulator (TCI) phase. The gapless node is also observed in our ARPES data and previous studies \cite{hosen2018observation}. Under the effect of SOC, the surface band spilts into a pair of bands with Rashba-type helical in-plane spin texture. The splitting can be seen in the inset of Fig. 1(d), and the spin texture is clearly shown with the constant energy contour near the gapless node in Fig. 1(f). The theoretical Rashba parameter estimated is 0.57 eV- $\angstrom$ with a momentum offset 0.0146  $\angstrom ^{-1}$. This indicates the slightly greater SOC strength in ZrGeTe compared to ZrSiS, where no Rashba effect is observed. Another isostructural compound HfSiS also exhibits Rashba splitting surface bands \cite{chen2017dirac} with Rashba parameter 3.1 eV-$\angstrom$. Note that by comparing theoretical calculation and QPI, we deduce the DFT calculation is 50 meV lower than the STM experiment, and thus we shift the theoretical data in Fig. \ref{fig:qpi}(k) - (m).
\newpage

\begin{figure}[h!] \centering \includegraphics[width=15cm,height=14cm]{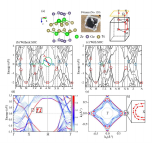}
	\caption{(a) ZrGeTe crystal structure and the corresponding Brillounin zone for bulk (black) and Te surface (orange). Bulk electronic band structure without (b) and with (c) SOC. The blue circles mark the nodes gapped by SOC, while the symmetry protected Dirac nodes are marked with red circles and form Dirac line nodes marked with red boxes. The green circle mark the type-II Dirac point without considering the spin degree of freedom, and the irreducible representation for corresponding bands are shown in the green inset. (d) 9-layer slab band sturcture with ratio of surface component indicated with color. The inset clearly shows the Rashba splitting. (e) 9-layer slab constant energy contour (CEC) at E = -0.2 eV, which is close to the gapless node at $\bar{X}$. The inset shows the in-plane helical spin texture for the Rashba splitting bands.  
	}
	\label{fig:lattice}
\end{figure}
\newpage

\section*{ARPES measurements}
\begin{figure}[h!] \centering \includegraphics[width=15cm,height=13cm]{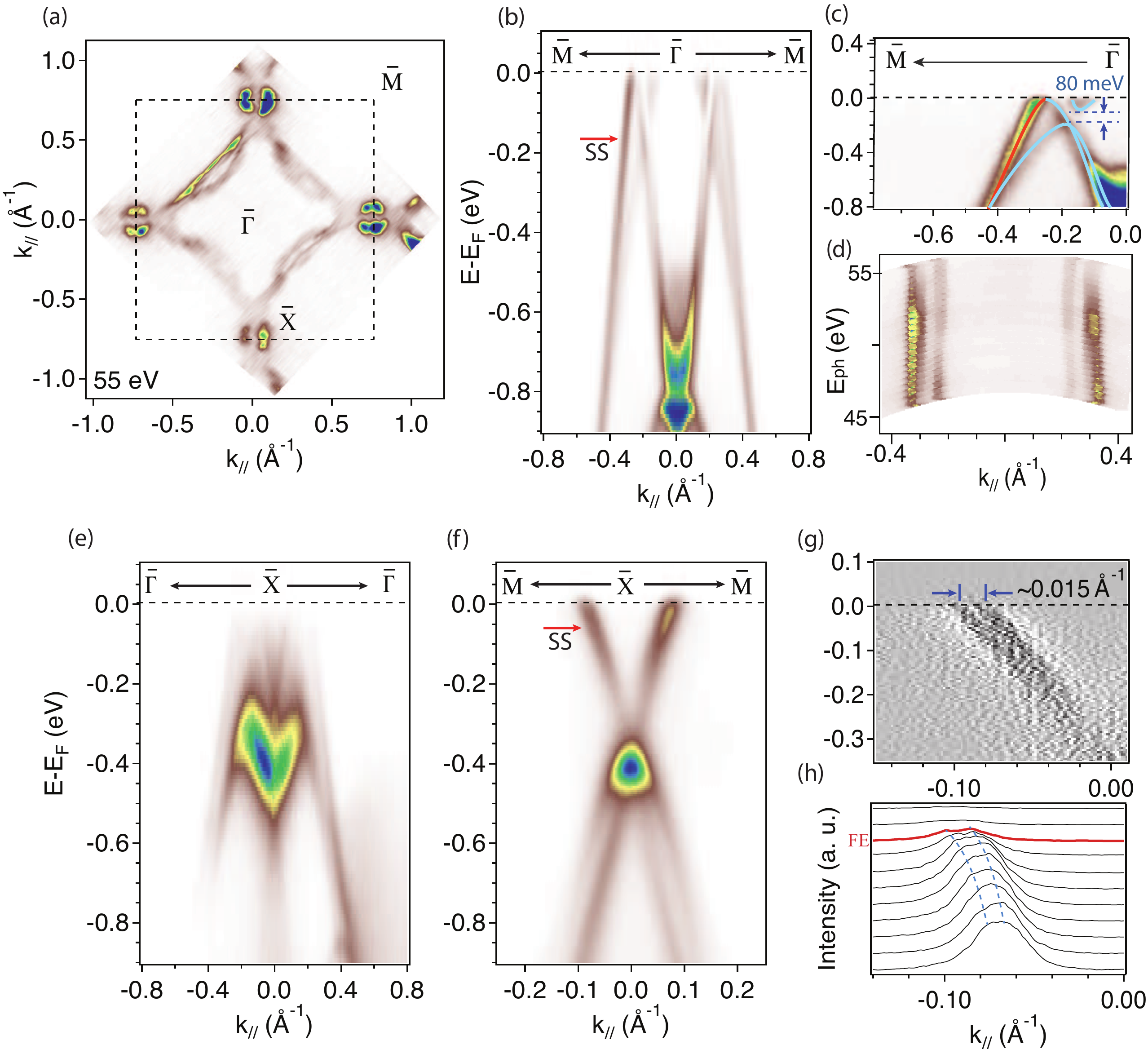}
	\caption{ (a) Fermi surface mapping in the k$_x$–k$_y$ plane measured at 80 K and E$_{ph}$= 55 eV. The intensity has been integrated with a window of 10 meV around Fermi level. Band Dispersion along high symmetry direction (b)(c)$\overline{M\Gamma M}$, (e)$\overline{\Gamma X \Gamma}$ and (f)$\overline{MXM}$. (c) A band gap of $\sim$80 meV is estimated by guides to the eye. The red and blue curves correspond to the SS component and the bulk band in our calculation, respectively. (d) The k$_z$ dependent intensity mapping at E$_F$ as a function of k$_{//}$ ($\overline{M\Gamma M}$ direction) and photon energy (45-55 eV). (g) Second derivative of band map of the floating SS around $\overline{X}$. (h) momentum distribution curves of (f) as a function of binding energy (E-E$_F$=0.1 eV to -0.4 eV). Dashed lines mark the double peak manifests the Rashba splitting. 
	}
	\label{fig:arpes}
\end{figure}

ARPES experiments were performed at beamline BL21B1 at National synchrotron Radiation Research center (NSRRC) in Taiwan with a Scienta R4000 analyzer at 80 K. The energy and angular resolutions were better than 30 meV and 0.2$^{\circ}$, respectively. The Fermi surface (FS) map measured at a 55 eV photon energy in Fig. \ref{fig:arpes}(a) reveals a large diamond shaped contour around $\overline{\Gamma}$. Circular Fermi pockets at $\overline{X}$ point are connected with delicate crescent like FS, and the diamond line gradually splits into the inner and outer part from the vertices to the $\overline{\Gamma M}$ cut. 

Fig. \ref{fig:arpes}(b)(e)(f) display the detailed electronic structure measured along the high symmetry directions $\overline{M\Gamma M}$, $\overline{\Gamma X \Gamma}$ and $\overline{MXM}$, respectively, which are qualitatively in good agreement with our calculation. Linearly dispersing bands originating from the surface state (SS) component are clearly observed at the outer side along $\overline{M \Gamma M}$ direction and around $\overline{X}$ point as a gapless V-shaped Dirac surface state (SS) (as indicated in Fig. \ref{fig:arpes}(b)(f)). 

It is worth noting that three-dimensional dispersion of the bulk bands is predicted by the theoretical calculation. However, k$_z$ dispersion along $\overline{M\Gamma M}$ exhibits a nearly straight feature. Such result suggests that the three-dimensionality in the bulk band is obscure (Fig. \ref{fig:arpes}(d)). A quasi-2D electronic structure in related ZrGe(Se,Te) materials is also reported \cite{hosen2018observation,nakamura2019evidence}. Nevertheless, in Fig. \ref{fig:arpes}(b)(c), the broadening of the inner pocket clearly reveals the three-dimensionality nature of the bulk band. A small gap ($\sim$80 meV, Fig. \ref{fig:arpes}(c)) slightly below Fermi level denotes the effect of SOC \cite{chen2017dirac,nakamura2019evidence}, agreeing with our calculation. The fact that this gap remains below Fermi level measured from 45 to 55 eV (which should cover at least one period $\overline{\Gamma}$-$\overline{Z}$) also implies a rather subtle three-dimensional electronic structure than theoretical prediction. 

The existence of Rashba splitting is another indication of SOC strength. Previous ARPES measurements on different compounds of this family also reported Rashba splitting on the SS at $\overline{X}$ \cite{chen2017dirac}. In our calculation, the Rashba splitting momentum offset in ZrGeTe is estimated to be 0.0146 $\angstrom^{-1}$. Fig. \ref{fig:arpes}(f) exhibits the floating gapless SS at $\overline{X}$. A trace of small splitting $\sim$0.015 $\angstrom^{-1}$ is observed by taking the second derivative of the ARPES intensity (see Fig. \ref{fig:arpes}(g)). In Fig. \ref{fig:arpes}(h), the existence of Rashba splitting is further evidenced by the set of double peaks in the momentum distribution curves. Our observation shows that the Rashba splitting in ZrSiS family is a generic feature even with modest SOC strength.

\newpage

\section*{STM imaging and QPI analysis}
\begin{figure}[h!] \centering \includegraphics[width=15cm,height=8.3cm]{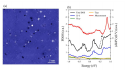}
	\caption{  STM topography and tunneling spectrum. (a) Topography of the cleaved ZrGeTe surface in a 40 x 40 nm$^2$ field of view (FOV). We mention this in Fig. \ref{fig:qpi}, where various defects or impurities exist as scattering centers for QPI (T=4.6K, V= 0.01V, I=100pA). (b) The measured tunneling spectrum and the calculated projected density of states (PDOS). The dip structure with corresponding arrow shows the tiny energy difference between STM experiments and the DFT calculation. The measured tunnelling spectrum (red) is shifted by 3 (a.u.) for clarity.
	}
	\label{fig:stm}
\end{figure}

To better understand the surface electronic structure across E$_F$, we further perform quasiparticle scattering interference (QPI) imaging on ZrGeTe by using a homemade $^3$Helium fridge STM system. We use the electrochemically etched tungsten tips cleaned by field emission and characterized on the gold surface. The ZrGeTe single crystal is cooled below 20K first and then cleaved in ultra-high vaccum (UHV) \textit{in-situ} before inserting to STM for measurements at T=4.6K. Fig. \ref{fig:stm}(a) shows a 40 x 40 nm$^2$ topography, which exhibits the 1 x 1 structure with lattice constant of a = 3.47 $\pm$ 0.29 $\angstrom$. Although it has been shown that ZrSiS can be cleaved to reveal surfaces with different termination\cite{su2018surface}, we have not observed terraces or different surfaces. We then investigate the local density of state (LDOS) by differential tunneling conductance (dI/dV) spectrum along with the calculated orbital-projected density of state (PDOS), as shown in Fig. \ref{fig:stm}(b). The tunneling spectrum shows the semi-metallic nature with lower density of states near Fermi level ($E_F$), and it agrees well with the calculated PDOS. The \textit{d} orbital of Zr dominates the DOS near $E_F$ and thus only Zr atoms are observed in STM topography in Fig. \ref{fig:stm}(a), which is similar to the case of ZrSiS \cite{su2018surface}. There are dip features in the calculated PDOS near E = 0.5 eV and in the tunneling spectrum near E = 0.6 eV respectively, which indicates the small energy shift between DFT calculation and STM at a high energy.

\begin{figure}[h!] \centering \includegraphics[width=15cm,height=12cm]{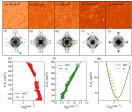}
	\caption{ QPI imaging on ZrGeTe surface. (a) - (e) Normalized differential conductance $dI/dV(\textbf{r},E)$ for five different energy layers taken on the 40 x 40 nm$^2$ field of view (FOV). (f) - (j) The corresponding Fourier transformed images $dI/dV(\textbf{q},E)$. The red circle in (f) corresponds to Bragg peak $2 \pi/a$ where $a$ is the lattice constant. The arrows and line segments with different colors indicate \textbf{q$_1$} - \textbf{q$_5$}. (k) - (m) The energy dispersion for \textbf{q$_1$} along $\overline{\Gamma M}$, \textbf{q$_2$} along $\overline{\Gamma X}$, and \textbf{q$_4$} along $\overline{\Gamma X}$ respectively. The black line shows the predicted QPI dispersion from DFT calculation.
	}
	\label{fig:qpi}
\end{figure}

\begin{figure}[h!] \centering \includegraphics[width=16cm,height=9cm]{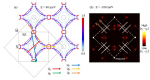}
	\caption{ (a) The calculated constant energy contours (CEC) at E=-40 meV. The ratio of surface component indicated with color is identical to the slab band structure in Fig. \ref{fig:lattice}(d). Five \textbf{q}-vectors \textbf{q$_1$} - \textbf{q$_5$} are shown with the corresponding bands resulting in the scattering. The solid line shows the first Brillouin zone (BZ), while the dashed line shows the \textit{P}4/\textit{nmm} nonsymmorphic reshaped BZ as discussed in \cite{zhen2018nonsym}. (b) The spin dependent joint density of states (JDOS) simulation of the diamond-shaped band resulting in \textbf{q$_1$} and \textbf{q$_5$} at E=-100 meV. The colorbar indicates the normalized dot product between spin components of the scattering states.  \textbf{q$_1$} and \textbf{q$_5$} are successfully reproduced. The dashed line indicates the boundary of \textbf{q$_5$} determined by the nonsymmorphic reshaped BZ as shown in subplot (a).
	}
	\label{fig:jdos}
\end{figure}

Next, we perform QPI imaging to resolve the surface electronic band structures. We measure the energy dependent differential tunneling conductance maps dI/dV(\textbf{r}, E=eV) at T=4.6K from E = -0.6 eV to E = 1.0 eV in the same field of view (FOV) of 40 x 40 nm$^2$ as Fig. \ref{fig:stm}(a). Fig. \ref{fig:qpi}(a)-(e) show dI/dV(\textbf{r}, E) conductance maps at five different energies, where defects induce complicated standing wave patterns. We then apply fast Fourier transforms (FFT) to the dI/dV(\textbf{r}, E) maps and obtain the correponding dI/dV(\textbf{q}, E) maps in Fig. \ref{fig:qpi}(f)-(j) respectively. Five prominent \textbf{q}-vectors with clear energy dispersion are observed in our QPI images, and we can identify their orgins by comparing with theoretical band structures. A detailed correspondence between the q-vectors and the calculated constant energy contour (CEC) at E = 40 meV is shown in Fig. \ref{fig:jdos}(a). \textbf{q$_1$} comes from intra-band scattering between the diamond-shaped bands surrounding $\overline{\Gamma}$, while \textbf{q$_2$} is originated from inter-band scattering between Rashba-splitting surface states near $\overline{X}$. The \textbf{q$_2$} scattering can only occur between two different spin channels with the same spin polarization by symmetry. The Umklapp scattering \textbf{q$_3$} is equal to \textbf{q$_2$} plus a reciprocal vector, and its relation with the \textit{P}4/\textit{nmm} nonsymmorphic Brillounin zone (BZ) has been discussed recently in ZrSiSe\cite{zhen2018nonsym}, where half of the \textbf{q$_3$} vectors are missing due to the zone folding. All three \textbf{q$_1$}, \textbf{q$_2$}, and \textbf{q$_3$} are hallmark scattering processes in the ZrSiS family that have been observed in previous STM experiments \cite{su2018surface,butler2017quasiparticle,lodge2017observation,zhen2018nonsym}.

Interestingly, we observe another two Umklapp processes that has yet been reported in ZrSiS family except in ZrGeTe. We deduce that \textbf{q$_4$} is the scattering between bulk-projected bands in the middle of $\overline{\Gamma X}$, and \textbf{q$_5$} comes from the diamond-shaped bands scattering crossing zone boundary. Similar to the relation between \textbf{q$_2$} and \textbf{q$_3$}, \textbf{q$_5$} equals to \textbf{q$_1$} add a reciprocal vector. The identification of \textbf{q$_5$} can be further supported by joint density of states (JDOS) simulation presented in Fig. \ref{fig:jdos}(b). Comparing the QPI imaging (Fig. \ref{fig:qpi}(i)) and the JDOS simulation (Fig. \ref{fig:jdos} (b)) both at E = -100 meV, the measured intensity of \textbf{q$_5$} vectors vanishes significantly over the dashed-line boundary in Fig. \ref{fig:jdos}(b), which indicates that \textbf{q$_5$} scattering vectors are still within the nonsymmorphic-reshaped boundary. In Fig. \ref{fig:qpi}(k) - (m), the energy dispersion of \textbf{q$_1$}, \textbf{q$_2$}, and \textbf{q$_4$} are extracted from our QPI results, and they agree well with our theoretical band structure. The observation of several Umklapp scattering on the ZrGeTe surface can stem from the close proximity of those bands to the zone boundary. Note that point defects or impurities with $C_{4}$ or $C_{2}$ symmetry are abundant on the surface, which in principle can produce band-selective scattering process, depending on their geometric or orbital characters\cite{butler2017quasiparticle}. Here, our QPI imaging is a result of averaging over all types of defects.

\section*{Summary}
In summary, we present a combined study on electronic properties of Dirac nodal line semimetal ZrGeTe by first principle calculation, ARPES measurements, and STM imaging. The calculated band structure shows several interesting features including slightly gapped type-II Dirac point, nonsymmorphic symmetry protected Dirac line nodes, and gapless surface bands with Rashba splitting. We further argue the existence of nontrivial TCI phase in ZrGeTe by symmetry indicators, which is responsible for the gapless floating surface bands at $\overline{X}$. Our ARPES measurements agree well with the calculation in band dispersion, and the Rashba splitting of 0.0146  $\angstrom ^{-1}$ is resolved. With joint density of states (JDOS) simulation, we can identify the origins of five QPI \textbf{q}-vectors, which all come from states with strong Zr-\textit{d} orbital contribution.  The two new Umklapp scattering vectors \textbf{q$_4$} and \textbf{q$_5$} can be explained through the recent discussion of Umklapp scattering in \textit{P}4/\textit{nmm} Brillounin zone \cite{zhen2018nonsym}. With the rich exotic quantum phenomena, ZrGeTe and related compounds can serve as a valuable platform for exploring Rashba effect, nonsymmorphic symmetry, and topological protection in TCI phase.

\section*{Acknowledgments}
T-MC is grateful to the financial support from Ministry of Science and Technology , Academia Sinica (AS-iMATE-108-11) and Kenda Foundation. L-PH is grateful to the support from Ministry of Science and Technology (105-2112-M-213 -003 -MY3). G-YG is grateful to the support from Ministry of Science and Technology (107-2112-M-002-012-MY3) and Far Eastern Y. Z. Hsu Science and Technology Memorial Foundation.

\newpage
\section*{Appendix}
\subsection{Computational details}
The electronic band structures are calculated based on density functional theory (DFT) with Vienna \textit{ab initio} simulation package (VASP) \cite{kresse1996efficient,kresse1993ab}. The pseudo-potentials used are GGA in Perdew-Burke-Ernzerhof (PBE) form \cite{perdew1996generalized} with projected-augmented-wave (PAW) method \cite{blochl1994projector}. The valence electrons of Zr, Ge, and Te considered are $4s^{2}4p^{6}4d^{2}5s^{2}$, $3d^{10}4s^{2}4p^{2}$, and $5s^{2}5p^{4}$ respectively. The plane wave cutoff energy is 400 eV. To study the surface states, we perform slab calculation with 9 layers supercell along (001) direction, with 20 Å vacuum layers. The termination plane in slab supercell is chosen as the cleavage plane in Fig. 1(a). In the self-consistent calculation, a 25x25x25 (40x40x1) $\Gamma$-centered Monkhorst-Pack k-mesh is used to sample the Brillouin zone for bulk (slab) calculation. 

\subsection{ZrGeTe crystal growth}
CVT growth method has been using Br$_2$, Cl$_2$ and I$_2$ as the vapor transport agents, which allows an effective and fast vapor transport to produce the necessary supersaturation for the expected final crystal. ZrGeTe single crystals grown in truncated square pyramid shape are shown in Figure 1(b). For the preparation of precursor powder materials, stoichiometric amount of 5 N pure elements in molar ratio of Zr : Ge : Te = 1 : 1 : 1.1 with 10 percents excess of Te was sealed into an evacuated quartz ampoule and heated for 2 days at 700 $^\circ$C. For the CVT method crystal growth, about 7.5 g of the pre-reacted ZrGeTe powder was placed together with variable amounts of transport agent (high purity) of either TeBr$_4$, TeCl$_4$, or solid I$_2$ at one end of the silica ampoule (35 cm long, ID/OD = 2.0/2.2 cm). To prevent oxygen contamination, all preparation steps before flame-sealing were carried out in an Ar gas filled glove box with oxygen and water level kept below $\sim$ 1 ppm. The loaded ampoule was evacuated and flame-sealed before loading into the tube furnace for the CVT growth. A temperature profile of gradient near 1.5 $^\circ$C/cm is maintained, where the pre-reacted material is held at 900 $^\circ$C and the growth end is maintained at 850 $^\circ$C, thin plate-like ZrGeTe single crystals were obtained with I$_2$ transport agent.  For the growth of thick crystals of  ZrGeTe in a truncated square pyramid shape, sizes up to 2.5 x 1.00 x 0.5 mm$^3$ and 2.0 x 2.5 x 1.00 mm$^3$ were obtained using transport agents of TeCl$_4$ and TeBr$_4$, respectively. The optimal temperature profile is kept at 980 $^\circ$C - 880 $^\circ$C with a small temperature gradient near 2.5 $^\circ$C/cm in about a week. It is found that the vapor pressure controlled at $\sim$ 6 mg of TeBr$_4$ per cm$^3$, $\sim$ 7 mg of TeCl$^4$ per cm$^3$, and $\sim$ 6 mg of solid I$_2$ per cm$^3$ would yield growth at high transport rate about $\sim$ 100, $\sim$ 120, and $\sim$ 155 mg per day, respectively. 

{}


\newpage
\section*{References}
\begin{thebibliography}{}

	

\bibitem{burkov2011topological} A. Burkov, M. Hook, and L. Balents, Phys. Rev. B \textbf{84}, 235126 (2011).

\bibitem{burkov2016topological} A. Burkov, Nat. Mater. \textbf{15}, 1145 (2016).

\bibitem{mellnik2014spin} A. Mellnik, J. Lee, A. Richardella, J. Grab, P. Mintun, M. H. Fischer, A. Vaezi, A. Manchon, E.-A. Kim, N. Samarth, and D.C. Ralph, Nature \textbf{511}, 449 (2014).

\bibitem{young2015dirac} S. M. Young and C. L. Kane, Phys. Rev. Lett. \textbf{115}, 126803 (2015).

\bibitem{schoop2016dirac} L. M. Schoop, M. N. Ali, C. Stra{\ss}er, A. Topp, A. Varykhalov, D. Marchenko, V. Duppel, S. S. Parkin, B. V. Lotsch, and C. R. Ast, Nat. Commun. \textbf{7}, 11696 (2016).

\bibitem{neupane2016observation} M. Neupane, I. Belopolski, M. M. Hosen, D. S. Sanchez, R. Sankar, M. Szlawska,  S. -Y. Xu, K. Dimitri, N. Dhakal, P. Maldonado, P. M. Oppeneer, D. Kaczorowski, F. Chou, M. Z. Hasan, and T. Durakiewicz, Phys. Rev. B \textbf{93}, 201104 (2016).

\bibitem{chen2017dirac} C. Chen, X. Xu, J. Jiang, S.-C. Wu, Y. Qi, L. Yang, M. Wang, Y. Sun, N. Schr{\"o}ter, H. Yang, L. M. Schoop, Y. Y. Lv, J. Zhou, Y. B. Chen, S. H. Yao, M. H. Lu, Y. F. Chen, C. Felser, B. H. Yan, Z. K. Liu, and Y. L. Chen, Phys. Rev. B \textbf{95}, 125126 (2017).

\bibitem{topp2016te} A. Topp, J. M. Lippmann, A. Varykhalov, V. Duppel, B. V. Lostch, C. R. Ast, and L. M. Schoop, New J. Phys. \textbf{18}, 125014 (2016).


\bibitem{hosen2017tune} M. M. Hosen, K. Dimitri, I. Belopolski, P. Maldonado, R. Sankar, N. Dhakal, G. Dhakal, T. Cole, P. M. Oppeneer, D. Kaczorowski, F.Chou, M.Z. Hasan, T. Durakiewicz, and M. Neupane, Phys. Rev. B \textbf{95}, 161101 (2017)

\bibitem{hosen2018observation} M. M. Hosen, K. Dimitri, A. Aperis, P. Maldonado, I. Belopolski, G. Dhakal, F. Kabir, C. Sims, M. Z. Hasan, D. Kaczorowski, T. Durakiewicz, P. M. Oppeneer, M. Neupane, Phys. Rev. B \textbf{97}, 121103 (2018).

\bibitem{nakamura2019evidence} T. Nakamura, S. Souma, Z. Wang, K. Yamauchi, D. Takane, H. Oinuma, K. Nakayama, K. Horiba, H. Kumigashira, T. Oguchi, T. Takahashi, Y. Ando, and T. Sato, Phys. Rev. B \textbf{99}, 245105 (2019).

\bibitem{ali2016butterfly} M. N. Ali, L. M. Schoop, C. Garg, J. M. Lippmann, E. Lara, B. Lotsch, and S. S. Parkin, Sci. Adv. \textbf{2}, e1601742 (2016).

\bibitem{su2018surface} C. C. Su, C. S. Li, T. C. Wang, S. Y. Guan, R. Sankar, F. C. Chou, C. S. Chang,W. L. Lee, G. Y. Guo, and T. M. Chuang, New J. Phys. \textbf{20}, 103025 (2018).

\bibitem{allah2019optical} J. Ebad-Allah, J. F. Afonso, M. Krottenmüller, J. Hu, Y. L. Zhu, Z. Q. Mao, J. Kuneš, and C. A. Kuntscher, Phys. Rev. B 9\textbf{9}, 125154 (2019).

\bibitem{topp2017float} A. Topp, R. Queiroz, A. Grüneis, L. Müchler, A. Rost, A. Varykhalov, D. Marchenko,M. Krivenkov, F. Rodolakis, J. McChesney, B. V. Lotsch, L. M. Schoop, C. R. Ast, Phys. Rev. X \textbf{7},041073 (2017).

\bibitem{butler2017quasiparticle} C. J. Butler, Y.-M. Wu, C.-R. Hsing, Y. Tseng, R. Sankar,  C. -M. Wei, F. -C. Chou, and M. -T. Lin, Phys. Rev. B \textbf{96},195125 (2017).

\bibitem{lodge2017observation} M. S. Lodge, G. Chang, C. Y. Huang, B. Singh, J. Hellerstedt, M. T. Edmonds, D. Kaczorowski, M. M. Hosen, M. Neupane, H. Lin, M. S. Fuhrer, B. Weber, M. Ishigami, Nano. Lett. \textbf{17}, 7213 (2017).

\bibitem{zhen2018nonsym}Z.  Zhu,  T.-R.  Chang,  C.-Y.  Huang,  H.  Pan,  X.-A.  Nie,  X.-Z.Wang, Z.-T. Jin, S.-Y. Xu, S.-M. Huang, D.-D. Guan, S. Wang,Y.-Y. Li, C. Liu, D. Qian, W. Ku, F. Song, H. Lin, H. Zheng, and J.-F. Jia, Nat. Commun. \textbf{9}, 4153 (2018).

\bibitem{song2018quant} Z. Song, T. Zhang, Z. Fang, and C. Fang, Nat. Commun. \textbf{9}, 3530 (2018).

\bibitem{zhang2019catalogue} T. Zhang, Y. Jiang, Z. Song, H. Huang, Y. He, Z. Fang, H.Weng, and C. Fang, Nature (London) \textbf{566}, 475 (2019).

\bibitem{verg2019catalogue} M. Vergniory, L. Elcoro, C. Felser, N. Regnault, B. A. Bernevig, and Z. Wang, Nature (London) \textbf{566}, 486 (2019).

\bibitem{kresse1996efficient} G. Kresse and J. Furthm{\"u}ller, Phys. Rev. B \textbf{54}, 11169 (1996).

\bibitem{kresse1993ab} G. Kresse and J. Hafner, Phys. Rev. B \textbf{47}, 558 (1993).

\bibitem{perdew1996generalized} J. P. Perdew, K. Burke, and M. Ernzerhof, Phys. Rev. Lett. \textbf{77}, 3865 (1996).

\bibitem{blochl1994projector} P. E. Bl{\"o}chl, Phys. Rev. B \textbf{50}, 17953 (1994).


\end{thebibliography}
\end{document}